\documentclass[english,showpacs,floatfix,preprint,12pt]{revtex4}

\usepackage[dvips]{graphicx}
\usepackage{longtable}
\usepackage{amsmath,amssymb}
\usepackage{dcolumn}
\usepackage{latexsym}
\usepackage{babel}
\usepackage[latin1]{inputenc}
\usepackage{color}
\usepackage[colorlinks]{hyperref}

\newcommand{\be}{\begin{equation}}
\newcommand{\ee}{\end{equation}}

\begin{document}
\title{Thermodynamical description of the interaction between holographic dark energy and dark matter}
\author{Bin Wang $^{1}$\footnote{E-mail address:
wangb@fudan.edu.cn}, Chi-Yong Lin $^{2}$\footnote{E-mail address:
lcyong@mail.ndhu.edu.tw}, Diego Pav\'{o}n$^{3}$\footnote{E-mail
address: diego.pavon@uab.es}, Elcio Abdalla $^{4}$\footnote{E-mail
address: eabdalla@fma.if.usp.br}} \affiliation{$^{1}$ Department
of Physics, Fudan University, 200433 Shanghai} \affiliation{$^{2}$
Department of Physics, National Dong Hwa University, Shoufeng, 974
Hualien} \affiliation{$^{3}$ Department of Physics, Autonomous
University of Barcelona, 08193 Bellaterra, Barcelona}
\affiliation{$^{4}$ Instituto de F\'{\i}sica, Universidade de
S\~ao Paulo, CP 66318, 05315-970, S\~ao Paulo}

\begin{abstract}
We present a thermodynamical description of the interaction
between holographic dark energy and dark matter. If holographic
dark energy and dark matter evolve separately, each of them
remains in thermodynamic equilibrium. A small interaction between
them may be viewed as a stable thermal fluctuation that brings a
logarithmic correction to the equilibrium entropy. From this
correction we obtain a physical expression for the interaction
which is consistent with phenomenological descriptions and passes
reasonably well the observational tests.
\end{abstract}


\maketitle

A variety of cosmological observations strongly suggest  that our
Universe is currently undergoing a phase of accelerated expansion
\cite{obs,wmap1,wmap3}, likely driven by some exotic component
called dark energy (DE) whose main feature is to possess a high
negative pressure -see however \cite{sarkar}. Nevertheless,
despite the mounting observational evidence, the nature and origin
of dark energy remains elusive and it has become a source of vivid
debate -see \cite{debate} and references therein.

Most discussions on DE rely on the assumption that it evolves
independently of other matter fields. One might argue that given
the unknown nature of both DE and dark matter (DM), an entirely
independent behavior of DE and DM is very special whereby it is
not unnatural to suppose that they interact. Studies on the
interaction (coupling) between DE and DM have been carried out in
\cite{diego}-\cite{amendola} and, in particular, it has been shown
that the coupling can alleviate the coincidence problem
\cite{diego,amendola}. Furthermore, it was argued that the
appropriate interaction between DE and DM can influence the
perturbation dynamics  and lowest multi-poles of the cosmic
microwave background (CMB) spectrum and account for the observed
CMB low $l$ suppression \cite{bin2,zim}. The strength of the
coupling could be as large as the fine structure constant
\cite{bin2,elcio}. Recently it was shown that such an interaction
could be inferred from the expansion history of the Universe, as
manifested in the supernova data together with CMB and large-scale
structure \cite{feng}.

In contrast to minimally coupled DE models, the coupling between
DE and DM not only influences the Universe expansion history but
also modifies the structure formation scenario through the
coupling to cold DM density fluctuations \cite{german,bean}.
Indeed, the growth of DM perturbations can be enhanced due to the
coupling \cite{bin2, das}, which can be used to explain the age
($\sim 2.1$ Gyr) of the old quasar APM0879+5255 observed at
redshift $z = 3.91$ \cite{bin2}. Further, lately it was suggested
that the dynamical equilibrium of collapsed structures would be
affected by the coupling of dark energy to dark matter in a way
that could be detected in the galaxy cluster Abell A586
\cite{orfeu}. Through the internal dynamics of galaxy clusters and
using reliable x-ray, weak lensing and optical data from 33 galaxy
clusters, a much tighter limit on the strength of the coupling
between DE and DM has been established \cite{raul}. Indeed, it was
shown that the coupling is small but positive which indicates that
DE may decay into DM. Nevertheless, albeit the interaction
hypothesis is gaining ground the observational limits on the
strength of the coupling remain weak \cite{guo}.

The interaction between DE and DM could be a major issue to be
confronted in studying the physics of DE. However, so long as the
nature of these two components remain unknown it will not be
possible to derive the precise form of the interaction  from first
principles. Therefore, one has to assume a specific coupling from
the outset \cite{das,amendola} or determine it  from
phenomenological requirements \cite{diego}. Nevertheless, attempts
to provide a Lagrangian description of the interaction have been
put forward. These comprise proposals that include the dependence
of the matter field on the scalar field \cite{shin} or express the
cosmological constant as a function of the trace of the
energy-momentum tensor \cite{pop};  at any rate, the exact form of
the dependence stays unspecified.

The main purpose of this Letter is to try to understand such a
coupling from thermodynamical considerations. The thermodynamics
of black hole physics \cite{hawking1} and  de Sitter space
\cite{hawking2} are well established. Recently, extensive analysis
found that the current data favors DE models with EoS very close
to $w_D = -1$. This suggests that the present evolution of the
universe is practically quasi-de Sitter. Therefore, one may assume
that some thermodynamical approach will also apply to eternally
accelerating quasi-de Sitter universes \cite{bousso}.

We shall assume that in the absence of a mutual  interaction both
DE and DM remain in their respective thermodynamic equilibrium
states, and that a small coupling between DE and DM may be viewed
as small stable fluctuations around equilibrium. (We say ``small"
because a large, or even moderate, coupling would substantially
deviate the model from the $\Lambda$CDM concordance model and
would be incompatible with observation \cite{german}). Some years
ago, Das {\it et al.} \cite{das2} showed that logarithmic
corrections to the equilibrium thermodynamic entropy arise in all
thermodynamic systems when stable fluctuations around equilibrium
are taken into account and that, in particular, it leads to
logarithmic corrections to the Bekenstein-Hawking formula for
black hole entropy. This idea was later applied to obtain an
evolution law for the cosmological constant \cite{bar}. We shall
present a thermodynamic description of the interaction between DE
and DM by building  a relation between the logarithmic entropy
correction and the interaction. Thus this derivation possesses a
solid physical foundation. Next, we will argue that our
thermodynamical interpretation of the interaction  is consistent
with phenomenological approaches and meet observational
constraints.

We shall focus on the DE model inspired by the holographic idea
that the energy within our horizon cannot exceed the mass of a
black hole of the same size \cite{cohen,li, gong}. The extension
of the holographic principle to a general cosmological setting was
first addressed by Fischler and Susskind \cite{2} and subsequently
got modified by many authors \cite{3}-\cite{7}. The idea of
holography is viewed as a real conceptual change in our thinking
about gravity \cite{8}. There have been a lot of attempts on
applying holography in the study of cosmology. It is interesting
to note that holography implies a possible value of the
cosmological constant in a large class of universes \cite{9}. In
an inhomogeneous cosmology holography was also realized as a
useful tool to select physically acceptable models \cite{6}. The
idea of holography has further been applied to the study of
inflation and gives possible upper limits to the number of e-folds
\cite{10}. Recently, holography has again been proved as an
effective way to investigate dark energy \cite{11,li}. Thus
holography seems a useful tool to investigate cosmology.

Following Li \cite{li}, we assume that the holographic dark energy
density is given by $\rho_D=3c^2/L^2$, where $c^2$ is a constant
of order unity and $L$ is an appropriate length scale which we
identify with the radius of the future event horizon,
\\
\begin{equation}\label{radius}
R_E =a\int_a^{\infty}\frac{dx}{H \, x^{2}} \, .
\end{equation}
\\
Here, $H \equiv \dot{a}/a$ is the Hubble function and $a$ the
scale factor of the Robertson-Walker metric.

The total energy density is $\rho=\rho_m+\rho_D$, where $\rho_m$
is the matter energy density and $\rho_D=3\, c^2/(8\pi R_E^2)$ is
the holographic DE energy density -we neglect radiation and
non-dark matter. If holographic DE and DM do not interact, their
energy densities satisfy separate conservation laws
\\
\begin{equation}\label{consvm1}
\dot{\rho}_m+3H\rho_m=0,
\end{equation}
\begin{equation}\label{consvd1}
\dot{\rho}_D+3H(1+w_D^0)\rho_D=0,
\end{equation}
where $w_D^0$ is the EoS of the holographic DE when it evolves
independently of DM. Introducing the dimensionless density
parameter for DE, $\Omega_D= 8 \pi \rho_D/(3H^2)$, the event
horizon radius can be written as \cite{li}
$R_E=c/(\sqrt{\Omega_D}H)$. Taking the derivative with respect to
$\ln a$ of last expression and resorting to Eq. (\ref{radius}) we
get
\begin{equation}\label{hprime}
\frac{H'}{H}=\frac{\sqrt{\Omega_D}}{c}-1-\frac{\Omega_D'}{2\Omega_D}
\, .
\end{equation}

Using  Friedmann's equation, $\Omega_D+\Omega_m=1$ and
(\ref{consvm1})-(\ref{hprime}), valid for spatially-flat
homogeneous isotropic cosmologies, we obtain for the EoS of the
holographic dark energy component the expression,
$w_D^0=-\frac{1}{3}-\frac{2\sqrt{\Omega_D}}{3c}$. Likewise, it
follows that the holographic DE evolution is governed by
\cite{bin1,li}
\\
\begin{equation}\label{omegaprime}
\Omega_D'=\Omega_D^2(1-\Omega_D)\left[\frac{1}{\Omega_D}
+\frac{2}{c\sqrt{\Omega_D}}\right]. 
\end{equation}

Equipped with these relationships, we can determine how much the
event horizon changes in one Hubble time,
\begin{equation}\label{hubbletime}
t_H\frac{\dot{R}_E}{R_E} =1-\frac{\sqrt{\Omega_D}}{c}\, ,
\end{equation}
\\
where $t_{H} \equiv H^{-1}$. Provided that $c = {\cal O}(1)$ -as
should be expected, see \cite{li}-  the event horizon will not
change significantly over one Hubble scale whereby the
thermodynamical description near equilibrium seems a reasonable
approach. It also follows from  Eq. (\ref{hubbletime}) that
$w_{D}^{0}
> -1 (<-1)$ if $\dot{R}_{E} >0 (<0)$, respectively.

The equilibrium entropy of the holographic DE component is related
to its energy and pressure by Gibbs' equation \cite{pavon3}
\\
\begin{equation}\label{gibbs}
TdS_D=dE_{D} +P_{D} \, dV.
\end{equation}
\\
Considering $V=4\pi R_E^3/3, E_{D} = \rho_{D} \, V= c^2 R_{E}/2$
and using the event horizon temperature $T=1/(2\pi R_{E})$, we get
\\
\begin{equation}\label{sd0}
dS_{D0}= \pi  c^2 \, (1+3w_D^0)R^0_E \, dR^0_E=- 2\pi
c\sqrt{\Omega^0_D}R^0_EdR^0_E,
\end{equation}
\\
for the holographic DE entropy when it is not coupled to DM. (A
zero superscript or subscript indicates absence of interaction).

However, when holographic DE and DM interact with each other, they
cannot remain in their respective equilibrium states. The effect
of the interaction may be assimilated to small stable fluctuations
around thermal equilibrium. It was shown that due to the
fluctuation, there is a leading logarithmic correction,
$S_1=-\frac{1}{2}\ln (CT^2)$ -with $C$ the heat capacity-, to the
thermodynamic entropy around equilibrium in all thermodynamical
systems \cite{das2}. In our case, the heat capacity of the DE can
be calculated as $C=T (\partial S_{D0}/\partial T)=-\pi
c^2(1+3w^0_D)(R^0_E)^2$, which is positive since for holographic
DE one has $1+3w_D^0<0$. Accordingly, the fluctuation is indeed
stable and the entropy correction reads
\\
\begin{equation}\label{sd1}
S_{D1}=-\frac{1}{2}\ln \left[-\frac{c^2}{4\pi}
(1+3w_D^0)\right]=-\frac{1}{2}\ln
\left[\frac{c}{2\pi}\sqrt{\Omega^0_D}\right].
\end{equation}

As mentioned above, we assume that this entropy correction is linked to
the DE-DM coupling. Thus, the total entropy of holographic DE
enclosed by the event horizon is $S_{D} = S_{D0} +S_{D1}$ and from
Gibbs' equation we get
\begin{equation}\label{1+3w}
1+3w_{D} = \frac{1}{c^2\pi R_E}\frac{dS}{dR_E}=\frac{1}{c^2\pi
R_E}\frac{dS_{D1}}{dR_E}-\frac{2\sqrt{\Omega^0_D}}{c}\frac{R^0_E}{R_E}
\frac{dR^0_E}{dR_E}\, ,
\end{equation}
\\
where $w_D$ denotes the EoS of holographic DE when it is coupled
to DM. If the interaction were turned off, the DE would return to
its equilibrium state and we would have that $w_D\rightarrow
w_D^0$ and $R_E\rightarrow R^0_E$.

When an interaction between holographic DE and DM exists, their
energy densities no longer satisfy independent conservation laws.
They obey instead
\begin{eqnarray}
\dot{\rho}_m+3H\rho_m &=&Q\quad ,\label{consvm2}\\
\dot{\rho}_D+3H(1+w_D)\rho_D&=&-Q\quad,\label{consvd2}
\end{eqnarray}
 $Q$ denotes the interaction term which is expected to be
derived from the entropy correction\footnote{In writing Eq.
(\ref{consvm2}) we have implicitly assumed that the DM continues
to be pressureless in spite of the presence of the interaction.
Obviously this is not strictly true, but since the interaction is
to be small the induced pressure will be much lower than
$\rho_{m}$ thereby we neglect it.}.

We first rewrite last two equations as
\begin{equation}\label{consvm3}
\Omega_m'+\frac{2H'}{H}\Omega_m+3\Omega_m=\frac{8\pi Q}{3H^3} \, ,
\end{equation}
\begin{equation}\label{consvd3}
\Omega_D'+\frac{2H'}{H}\Omega_D+3(1+w_D)\Omega_D=-\frac{8\pi Q}{3H^3},
\end{equation}
\\
and insert (\ref{hprime}) into (\ref{consvd3}) to get
\\
\begin{equation}\label{1+3wagain}
1+3w_D=-\frac{2\sqrt{\Omega_D}}{c}-\frac{8\pi Q}{3H^3\Omega_D}.
\end{equation}
\\
Then,  comparing last expression with Eq. (\ref{1+3w}), we obtain
\\
\begin{equation}\label{Q1}
\frac{8\pi
Q}{9H^3}=\frac{\Omega_D}{3}\left[-\frac{2\sqrt{\Omega_D}}{c}+
\frac{2\sqrt{\Omega^0_D}}{c}\frac{R^0_E}{R_E}\frac{dR^0_E}{dR_E}\right]-\frac{1}
{\pi c^2 R_E}\frac{\Omega_D}{3}\frac{dS_1}{dR_E}
\end{equation}
\\
for the interaction term, $Q$.

>From (\ref{sd1}) the evolution of $S_{D1}$ appearing in the last
equation can be written as
\\
\begin{equation}\label{evolsd1}
\frac{dS_1}{dR_E}=-\frac{H}{(c/\sqrt{\Omega_D})-1}\frac{(\Omega^0_D)'}{4\Omega^0_D}\,
,
\end{equation}
\\
where we have made use of  $R_E=c/(H\sqrt{\Omega_D})$.

We, thus, have built a relation between the DE-DM coupling and the
correction, $S_{D1}$, to the equilibrium entropy.

To see how the above expression for $Q$ (equations (\ref{Q1}) and
(\ref{evolsd1})) fares  when contrasted with observation let us
compare it with the interaction term \cite{diego}
\\
\begin{equation} \label{Q2}
Q=3b^2 \, H\, (\rho_{m} +\rho_{D})\, ,
\end{equation}
\\
where $b^2$ is a coupling constant, introduced on phenomenological
grounds  to alleviate the coincidence problem \cite{paul}.
However, before doing that let us provide a rationale for
(\ref{Q2}).

The  right hand side of (\ref{consvm2}) and (\ref{consvd2}), i.e.,
$Q$ and $-Q$,  must be functions of the energy densities
multiplied by a quantity with units of inverse of time. For the
latter the obvious choice is the Hubble factor $H$, so we have
that $Q = Q( H\rho_{m}\, , H\rho_{D})$. By power law expanding $Q$
and retaining just the first term we get $Q \simeq \lambda_{m}\, H
\rho_{m} + \lambda_{D}\, H \rho_{D}$. To facilitate comparison of
the resulting model with observation it is expedient to eliminate
one the two $\lambda$ parameters. Thus we set $\lambda_{m} =
\lambda_{D} = 3 b^{2}$ and arrive to Eq. (\ref{Q2}). The simpler
choice, $\lambda_{m}=0$ would not yield a constant dark matter to
dark energy ratio at late times. Clearly, the term $3b^{2}$
measures to what extent the decay rate of DE into DM differs from
the expansion rate of the Universe and also gauges the intensity
of the coupling. The lower $b^2$, the closer the evolution of the
Universe  to a non-interacting model is. It should be emphasized
that this phenomenological description has proven viable when
contrasted with observations, i.e., SNIa, CMB, large scale
structure, $H(z)$, and age constraints \cite{bin1,bin2,feng,guo},
and recently in galaxy clusters \cite{orfeu,raul}.

So, to carry out the said comparison we set $b^2=\frac{8\pi
Q}{9H^3}$. Accordingly, $b^{2}$ is no longer a constant but a
variable parameter that evolves according to
\\
\begin{equation}\label{b2}
b^2=\frac{8\pi Q}{9H^3}=
\frac{2\Omega_D^{3/2}}{3c}\left[-1+\frac{H^2\sqrt{\Omega_D}}{(H^0)^2\sqrt{\Omega^0_D}}\frac{\sqrt{\Omega^0_D}/c-1}
{\sqrt{\Omega_D}/c-1}\right]+\frac{1}{12 \pi
c^2}\frac{H^2}{c/\sqrt{\Omega_D}(c/\sqrt{\Omega_D}-1)}\frac{\Omega_D}{\Omega^0_D}\,
(\Omega^0_D)' \, .
\end{equation}
\\
Using Friedmann's equation as well as (\ref{hprime}), equation
(\ref{consvd3}) can be recast as
\\
\begin{equation}\label{omegaprimeagain}
\frac{\Omega_D'}{\Omega_D}+(\Omega_D-1)+\frac{2\sqrt{\Omega_D}}{c}(\Omega_D-1)=-\frac{8Q}{3H^3}=-3b^2.
\end{equation}

\begin{figure}[t] \label{fig1}
\begin{center}
\includegraphics[width=12cm,height=6cm,angle=0]{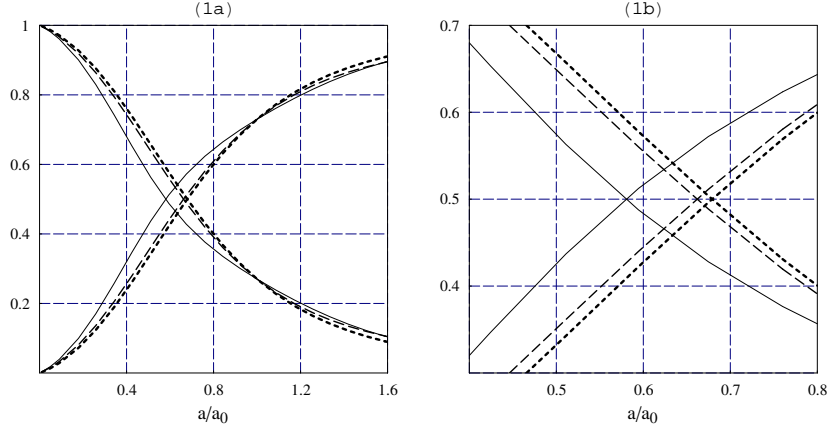}
\end{center}
\caption{ Evolutions of $\Omega_D$ and $\Omega_m$ with and without
interaction. Lines showing values increasing with $a$ is
$\Omega_D$, and the decreasing lines are for $\Omega_m$. The
solid, dotted, and dashed lines correspond to our scenario, the
holographic model without interaction, and the phenomenological
interacting model with $b^2=0.06$, respectively. }
\end{figure}

\begin{figure}[t] \label{fig2}
\begin{center}
\includegraphics[width=16cm,height=5cm,angle=0]{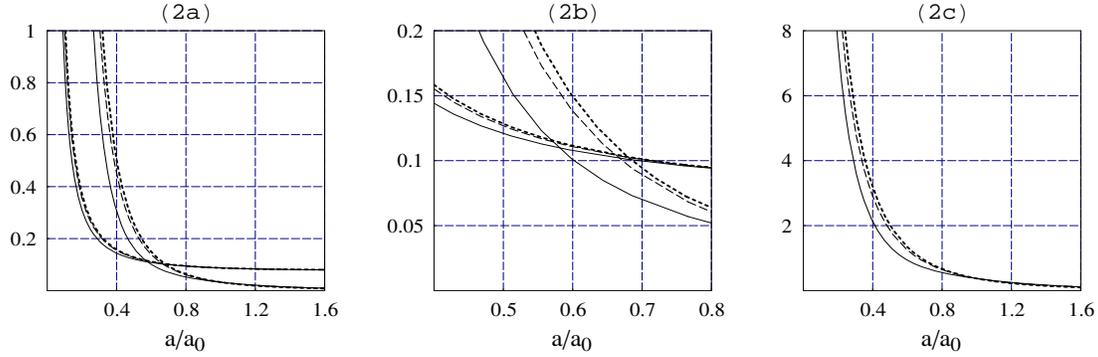}
\end{center}
\caption{Evolutions of $\rho_D$ and $\rho_m$ with and without
interaction. Before the crossing point, lines on the left are for
$\rho_D$, other bunch of lines are for $\rho_m$. The solid,
dotted, and dashed lines correspond to our scenario, the
holographic model without interaction, and the phenomenological
interacting model with $b^2=0.06$, respectively.}
\end{figure}

\begin{figure}[t] \label{fig3}
\begin{center}
\includegraphics[width=6cm,height=6cm,angle=0]{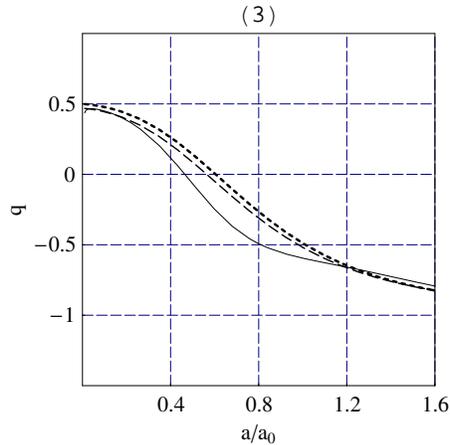}
\end{center}
\caption{ Dependence of the deceleration parameter, $q =
-\ddot{a}/(aH^2)$, on the interaction. The solid, dotted, and
dashed lines correspond to our scenario, the holographic model
without interaction, and the phenomenological interacting model
with $b^2=0.06$, respectively.}
\end{figure}

With the help of Eqs. (\ref{b2}), (\ref{omegaprimeagain}) and
(\ref{hprime}), we are in position to discuss the dependence of
the evolution of holographic DE in terms of the coupling to DM. In
the numerical calculations, we set $c=1$. From Fig.1 we learn that
because of the interaction between holographic DE and DM,
$\Omega_D$ increases faster, and from Figs.2a and 2b that
$\rho_{D}$ and $\rho_{m}$ follow each other and that the instant
at which $\rho_D=\rho_m$ occurs earlier than in the
non-interacting case. The latter feature is more clearly
appreciated in Fig. 2c where the dependence of the ratio $r \equiv
\rho_m/\rho_D$ with the scale factor is depicted.  The said ratio
decreases monotonously with expansion and it varies very slowly at
the present era. Compared with the noninteracting case, we find
that currently  $r$ decreases slower when there  is interaction.
This means, on the one hand, that the coincidence problem gets
substantially alleviated and, on the other hand, that in the
recent history of the Universe DE is decaying into DM. This is
consistent with phenomenological interacting models \cite{diego},
\cite{bin2}. The different evolution of the DM due to its
interaction with the DE gives rise to a different expansion
history of the Universe and a different evolution of the matter
density perturbations which alters the standard structure
formation scenario as the latter assumes $\rho_{m} \propto
a^{-3}$. In \cite{german,bin2} the matter density perturbations in
interacting models were investigated and in \cite{bin2} the impact
of the interaction on the DM perturbations was used to explain why
it is possible, as recently observed \cite{old}, for the old
quasar APM0879+5255 to exist already at the early stages of the
Universe (at $z= 3.91$). As a comparison, in Figs. 1 and 2 we have
also included  the phenomenological interaction case with constant
coupling, $b^2$ (dashed line). It is seen that the results
obtained for the evolution of holographic DE and DM using the
phenomenological model and using the interaction derived from the
thermodynamical consideration are consistent with each other.

Clearly, the interaction must affect the acceleration history of
the Universe. Figure 3 depicts the dependence of the deceleration
parameter, $q = - \ddot{a}/(aH^2)$, on the coupling. It is seen
that the interaction shifts the beginning of the acceleration to
earlier times; a result previously obtained by several authors
\cite{bin1}, \cite{diego}, \cite{das}, \cite{luca}.

\begin{figure}[t] \label{fig4}
\begin{center}
\includegraphics[width=6cm,height=6cm,angle=0]{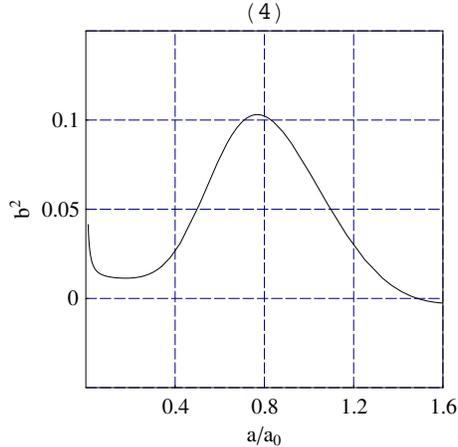}
\end{center}
\caption{ The corresponding coupling $b^2$ in our scenario by
comparing it with the simple phenomenological model.}
\end{figure}

Now we test this scenario for the interaction between holographic
DE and DM by using some observational results. For the comparison
with the phenomenological interacting model, in our scenario the
coupling between holographic DE and DM can be expressed as a
counterpart of $b$ as in the phenomenological interaction form.
Now the coupling is not longer a constant but  a time-dependent
parameter. Its evolution is depicted in Fig. 4.  During an ample
period, the effective coupling, $b$, remains small and positive,
indicating that holographic DE could be decaying into DM. In fact,
$b^2$ lies within the region of the golden supernova data fitting
result $b^2=0.00^{+0.11}_{-0.00}$ \cite{bin1} and the observed CMB
low $l$ data constraint \cite{bin2}. In Ref. \cite{bin2} it was
investigated whether this model satisfies the current Universe age
constraints and allows a considerably older universe at high
redshift to be compatible with the existence of some old objects
such as the old quasar APM0879+5255  at redshift $z=3.91$
\cite{old}. Its age, at that redshift,  was estimated as $t_g=2.1
Gyr$. Using the present WMAP data on the Hubble parameter, $H_0
=73.4^{+2.8}_{-3.8}$ \cite{wmap3}, the dimensionless age of the
quasar $T_g=H_0 \, t_g$ is seen to lie in the interval $0.148\leq
T_g\leq 0.162$. In our scenario, it is easy to realize that the
age of the Universe at $z=3.91$ was $T_z
=\int_{3.91}^{\infty}(1+z)^{-1}H^{-1}dz= 0.152$. This is to say,
at that redshift the Universe was old enough to accommodate the
existence of this old quasar. These results show that our
interacting DE scenario is compatible with observations.

In summary, from thermodynamical considerations we derived an
expression for the interaction between holographic DE and DM.  We
assumed that in the absence of a DE-DM coupling these two
components remain in separate thermal equilibrium and that the
presence of a small coupling between them can be described  as
stable fluctuations around equilibrium. Then, resorting to the
logarithmic correction to the equilibrium entropy \cite{das2} we
arrived to an expression for the interaction term, namely, Eq.
(\ref{Q1}) together with (\ref{evolsd1}). By comparing it with
phenomenological proposals, Eq.(\ref{Q2}), we concluded that this
scenario is compatible with the golden SN Ia data, small $l$ CMB
data and age constraints at different redshifts. The study here is
limited to the particular case of the holographic model. Our
argument may well not apply  to the Chaplygin gas model and its
generalizations \cite{gas}, since the admixture and interaction of
the DE and DM in these models does not imply any sort of entropy.
However, it would be interesting to generalize our work to  models
where DE and DM are not intrinsically mixed.

\begin{acknowledgments}
We thank the anonymous referee for constructive comments. This
work was partially supported by the NNSF of China, Shanghai
Education Commission, Science and Technology Commission; the
National Science Council ROC under the Grant
NSC-95-2112-M-259-003; FAPESP and CNPQ of Brazil; the Spanish
Ministry of Education and Science under Grant FIS
2006-12296-C02-01, and the ``Direcci\'{o} General de Recerca de
Catalunya" under Grant 2005 SGR 000 87.
\end{acknowledgments}


\end{document}